\documentclass{emulateapj}
\usepackage{epsfig}
\usepackage{color}
\usepackage{amsmath}
\usepackage{graphicx}
\usepackage[usenames,dvipsnames,x11names]{xcolor}
\usepackage{natbib}

\shorttitle{Local perturbations of the upper layers of a Sun-like star}
\shortauthors{Brito \& Lopes}

\begin{document}


\title{Local perturbations of the upper layers of a sun-like star:\\
  the impact on the acoustic oscillation spectrum}


\author{Ana Brito\altaffilmark{1,2,4}, Il\'\i dio Lopes\altaffilmark{1,3,5}}

\altaffiltext{1}{Centro Multidisciplinar de Astrof\'{\i}sica, Instituto Superior T\'ecnico, 
Universidade Tecnica de Lisboa , Av. Rovisco Pais, 1049-001 Lisboa, Portugal} 
\altaffiltext{2}{Departamento de Matem\'atica, Instituto Superior de Gest\~ao, Lisboa - Portugal}  
\altaffiltext{3}{Departamento de F\'\i sica, Escola de Ciencia e Tecnologia, 
Universidade de \'Evora, Col\'egio Luis Ant\'onio Verney, 7002-554 \'Evora - Portugal} 
\altaffiltext{4}{E-mail: ana.brito@ist.utl.pt}  
\altaffiltext{5}{E-mail: ilidio.lopes@ist.utl.pt} 


\begin{abstract}

In the last decade the quality and the amount of observational asteroseismic data that has been made available by space based missions had a tremendous upgrowth. The determination of asteroseismic parameters to estimate the fundamental physical processes occurring in stars' interiors, can be done today in a way that has never been possible before. In this work we propose to compute the seismic observable $\beta$, which is a proxy of the phase shift of the acoustic modes propagating in the envelope of the Sun-like stars. This seismic parameter $\beta$ can be used to identify rapid variation regions usually known as glitches. We show that a small variation in the structure, smaller than 1\% in the sound speed, produces a glitch in the acoustic potential that could explain the oscillatory character of $\beta$. This method allows us to determine the location and the thickness of the glitch with precision. We applied this idea to the Sun-like star $\alpha$ Centauri A and found a glitch located at approximately $0.94\,R$ (1400 s) and with a thickness of 0.2\% of the stars' radius. This is fully consistent with the data and also validates other seismic tests. 

\end{abstract}


\keywords{stars: interiors stars: oscillations stars: solar-type}

\maketitle

%
%
\section{Introduction\label{sec-intro}}
Asteroseismology was born as a natural extension of helioseismology (for a recent review see \citealt{2013arXiv1303.1957C}).
The success of helioseismology results from the ability to extract information from the Sun's interior through 
accurately measured frequencies of more than 7000 acoustic oscillation modes 
(e.g., \citealt{2011LNP...832....3K}; \citealt{2011RPPh...74h6901T}; \citealt{2012RAA....12.1107T}).
The development of this discipline was instrumental to improve the physics of  the solar standard model 
that led to the resolution of the solar neutrino problem (e.g., \citealt{2013ARA&A..51...21H}; \citealt{1993ApJ...408..347T}).
Similarly, 
asteroseismology  is expected to extract information from the interiors of stars at different stages of their evolution, 
through the study of  their observed  oscillation spectrum.
However,  in this case,  only global,  low degree modes are expected to be detected, due to observational limitations. In the next decade, an unprecedented progress is expected in this research field,
in part due to the large amount of data that has been made available by the missions {\it{CoRoT}} \citep{2008Sci...322..558M} and {\it{Kepler}} \citep{2010PASP..122..131G}. Up until now these missions have already detected oscillations in more than 500 stars \citep{2011Sci...332..213C}. These seismic data are obtained with an unprecedented high precision, which  
opens the possibility of studying new and old pulsating stars, as well as to test the theories of stellar evolution 
and probe for the existence of new physics inside stars \citep{2010Sci...330..462L, 2010ApJ...722L..95L, 2013ApJ...765L..21C}.  

A primordial target in asteroseismic studies is the class of  Sun-like stars, since these  stars are expected to have an internal structure and acoustic spectrum of oscillations 
identical to the Sun. There are a few Sun-like pulsations  for which a few dozen global acoustic modes 
were successfully identified, typically acoustic modes with a degree smaller than $4$. 
In fact, these global oscillations are particularly interesting to study because 
these modes travel from the surface  up to the core of the star, 
and thus these acoustic modes carry information  about the entire structure of the star. 
Moreover, the interest of studying this class of stars is twofold: 
(1) to better understand the physics behind stellar evolution in general;  
and (2) to better describe the physics occurring inside our own Sun. 
Like the Sun, in many Sun-like stars, the intricate surface structure where there are strong interactions among the magnetic fields, the turbulent convection and the rotation, as well the steady flows of meridional circulation, allow the possibility of local glitches that can produce visible perturbations in the spectrum of stellar oscillations.  It is our intent to test this theoretical scenario thoroughly and without the loss of generality.  We have focused the work in a Sun-like star which is the well known $\alpha$ Centauri A. We may have found some evidence of a perturbation of this type occurring in the oscillation spectrum of this star. 

\begin{figure*}
\centering
\plottwo{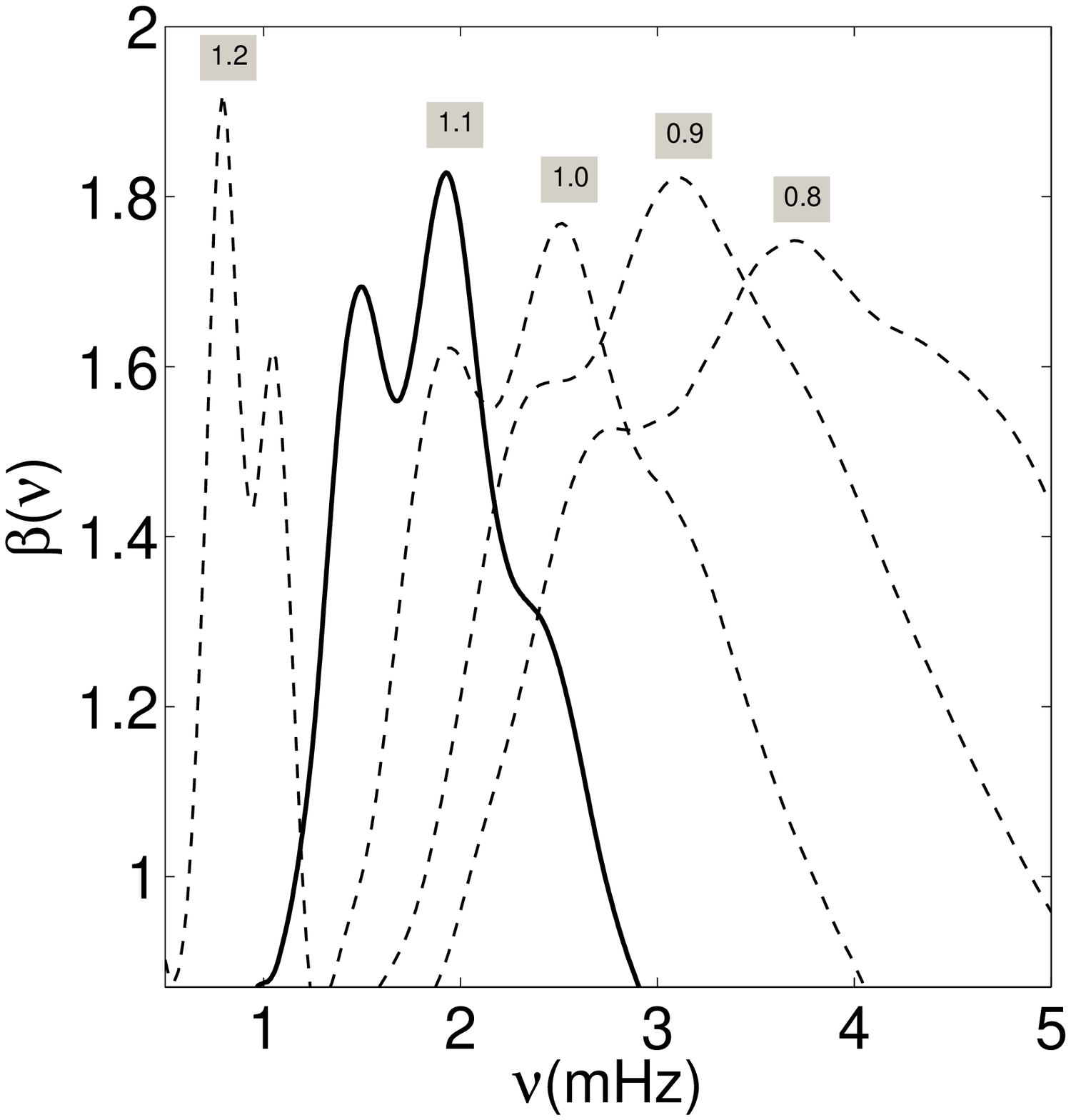}{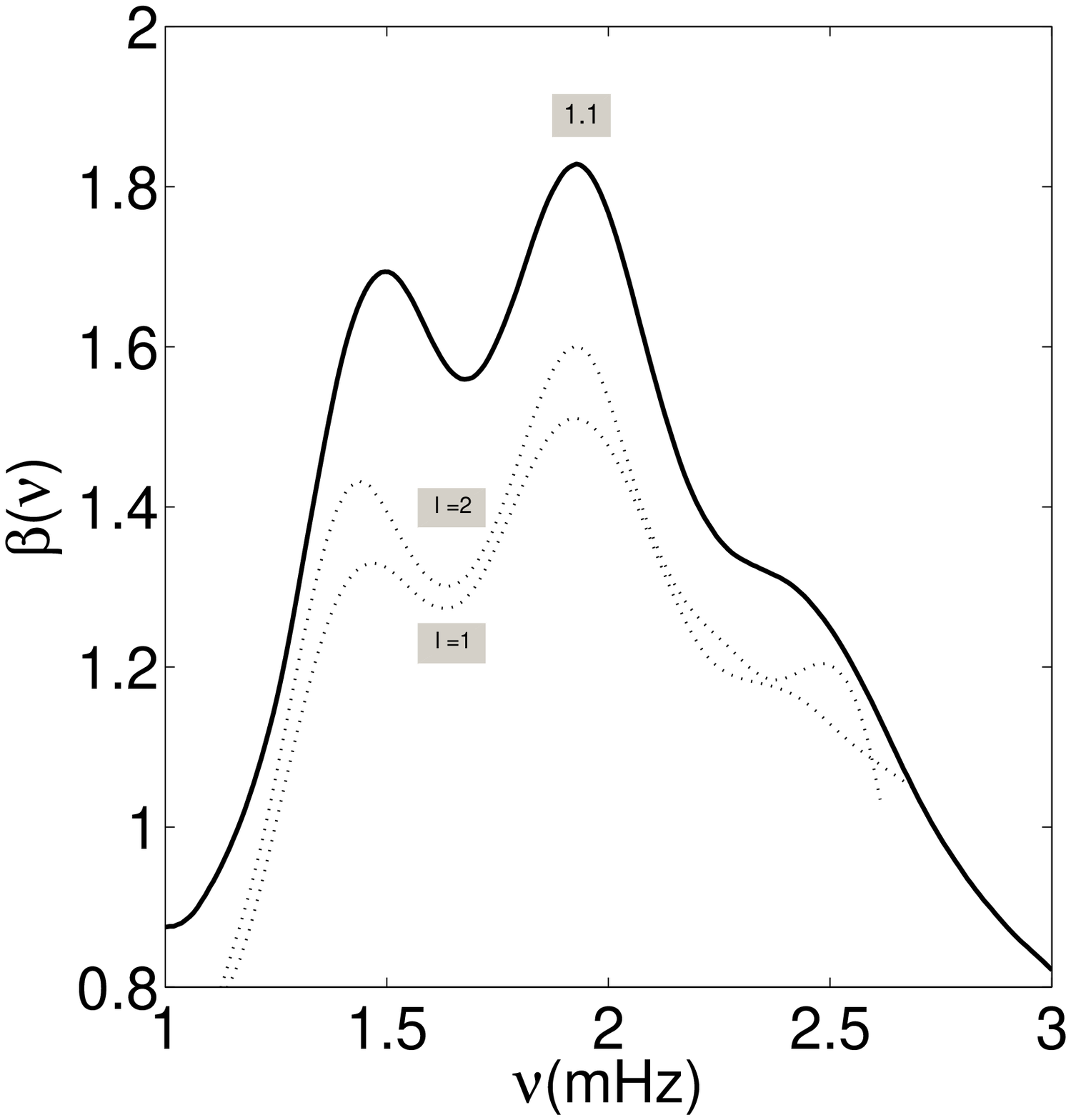}
\caption{Theoretical seismic signature $\beta$. Left: the continuous curve was obtained by numerical differentiation from the surface acoustic potential of a model of a Sun-like star with $1.1\, M_{\odot}$. Dashed lines show the theoretical $\beta(\nu)$ for a set of Sun-like stars with masses $0.8\, M_{\odot}, 0.9\, M_{\odot}, 1.0\, M_{\odot}$ and $1.2\, M_{\odot}$. Right: $\beta(\nu)$ was also calculated from a theoretical frequency table (dotted lines) for modes with the degrees $l=1$ and $l=2$. }
\end{figure*}

In this work, we discuss the structure of the upper layers of a Sun-like star, as it is possible to be inferred from   
the most complete observational table of frequencies. In particular, we apply the acoustic phase shift method \citep[see][for detaills]{1991SvA....35..400V, 1994MNRAS.268..880R, 2001MNRAS.322..473L, 2001MNRAS.321..615L} to probe the physical mechanisms operating in these upper layers. In this Sun-like star, we show that the existence 
of a rapid variation (RV) layer in the external region of the star will produce a significant amount of scattering of acoustic waves. This relatively thin layer 
is quite likely produced by an RV in the sound speed, which is similar to variations of sound speed that can occur in transition layers such as the well known examples in the Sun of the bottom  of the convection zone and the He II ionization zone. The seismic diagnostic is done by means
of the seismic parameter, which allows us to infer the scattering of acoustic waves in the  
surface of the star using  the available table of observational frequencies.

In Section 2, different methods are shown to compute
the scattering of acoustic waves in the upper layer of the star, and the impact of a RV of sound speed produced in the acoustic potential and consequently in the acoustic phase shift. 
In Section 3 we apply this method to a toy model of a Sun-like star, $\alpha$ Centauri A, and discuss the seismic diagnostics obtained from the observational data available for this star.  
A comparison is made between the theoretical data and the observational data. This comparison suggests that the internal structure of the stars could be quite different from what is predicted by the theoretical evolutionary models. It specifically indicates the presence of a thin layer where the sound speed undergoes an abrupt variation. Finally, in Section 4, we present  a summary and the conclusions of this work.

\section{The upper layers of a sun-like star: the seismic diagnostic}  

The asymptotic theory of  adiabatic, non-radial oscillations has been used and developed in the past, based on the fact that
the  oscillation modes have wavelengths shorter than the scale of local variations of the background state.
However, in the regions near the stellar surface, this approximation is no longer valid. In fact, the thermal and dynamical time scales 
in the stellar surface are of the same order of magnitude, contrary to what happens in the interior of a Sun-like star, for which the former is much larger than the latter. 
Consequently, the classical asymptotic theory gives a poor account 
of the propagation of waves in these more external layers. Strictly speaking, this means that the oscillations are no longer adiabatic, 
and an important exchange of energy occurs between the propagating wave and the background state.  To solve this problem,  \citet{1991SvA....35..400V} has introduced a frequency dependency on phase shift to take into account the scattering of waves
in these external layers. This phase shift frequency dependency is an obvious generalization of the reflection condition that occurs for all waves when reflected in a simple stellar isothermal atmosphere~\citep{1989nos..book.....U, 1993afd..conf..399G}. The phase shift $\alpha$ rather
than being a constant value for  all acoustic modes, such  as $\alpha=\mu/2$ where $\mu$ is the polytropic index 
of an isothermal atmosphere of a star like the Sun for which $\mu=3.0$, now becomes dependent on the frequency of the mode, such as $\alpha\equiv\alpha(\nu)$ where $\nu$ is the frequency of the mode of oscillation.   Furthermore, it was shown that  this phase shift can be partly inferred  by means of  the seismic observable $\beta (\nu)$, a proxy of the phase shift $\alpha  (\nu)$ that is computed either from the  phase shift scattering occurring in the  stellar envelope,
or from a table of frequencies of the modes that propagate through the same region. 
Several methods were developed to determine the scattering of 
acoustic waves in the outer layers of the Sun and stars.
A detailed discussion can be found in the literature \citep[e.g.,][]{1989SvAL...15...27B,1991SvA....35..400V,1992MNRAS.257...62C,1994MNRAS.268..880R,1994A&A...290..845L,2001MNRAS.322..473L}.

\subsection{Computing the Phase Shift from a Stellar Envelope}

\begin{figure*}
\centering
\plottwo{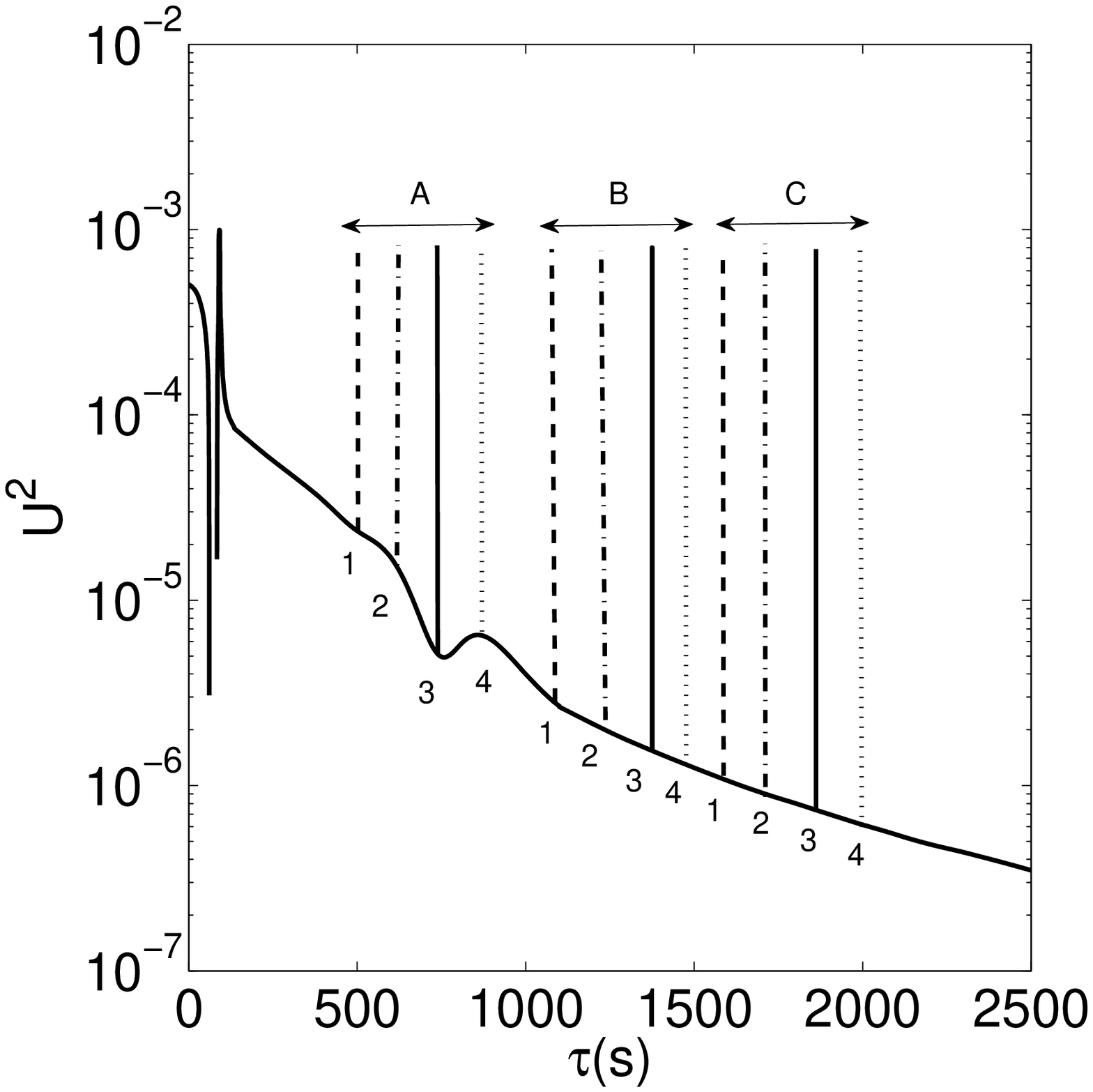}{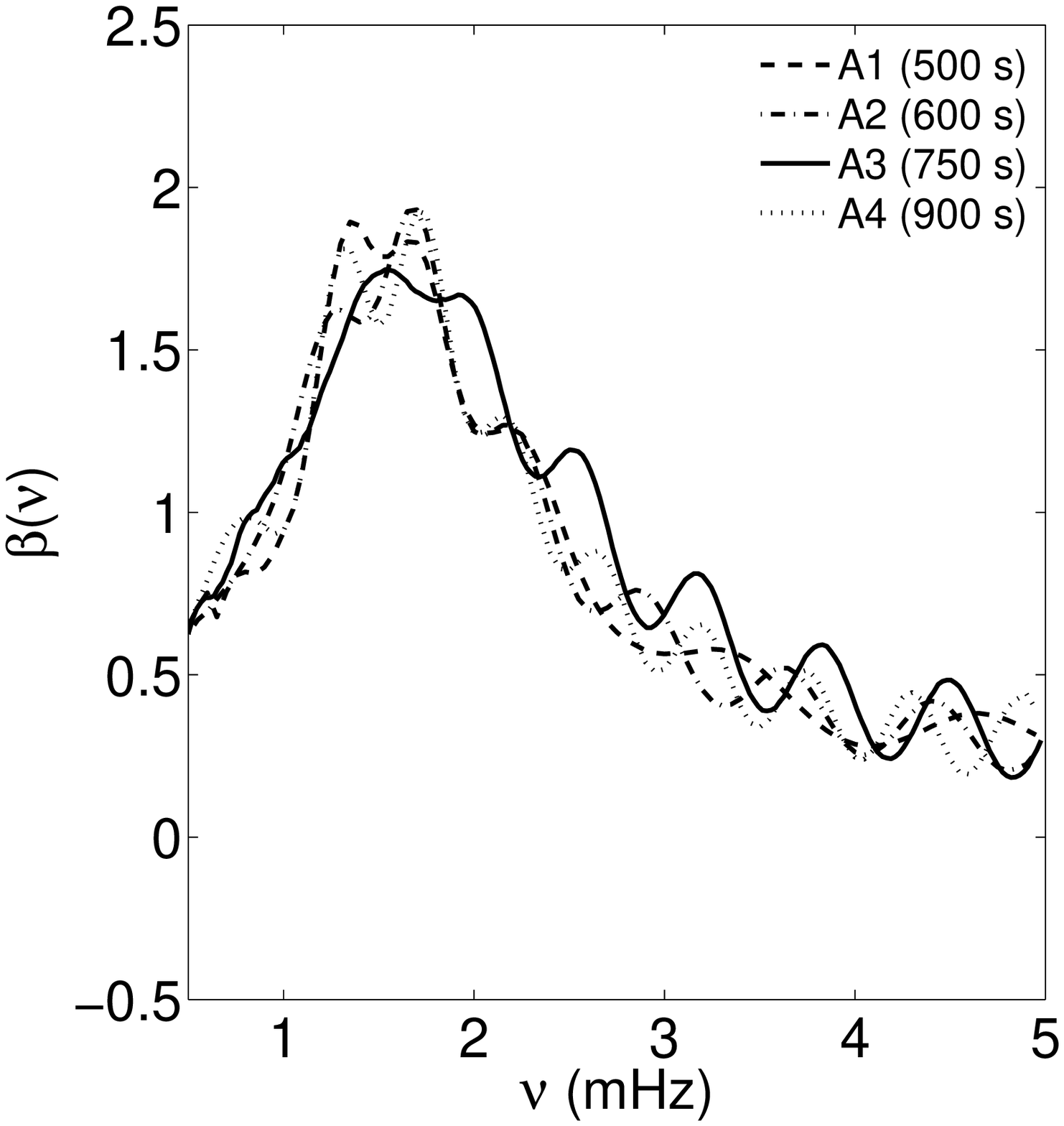}
\plottwo{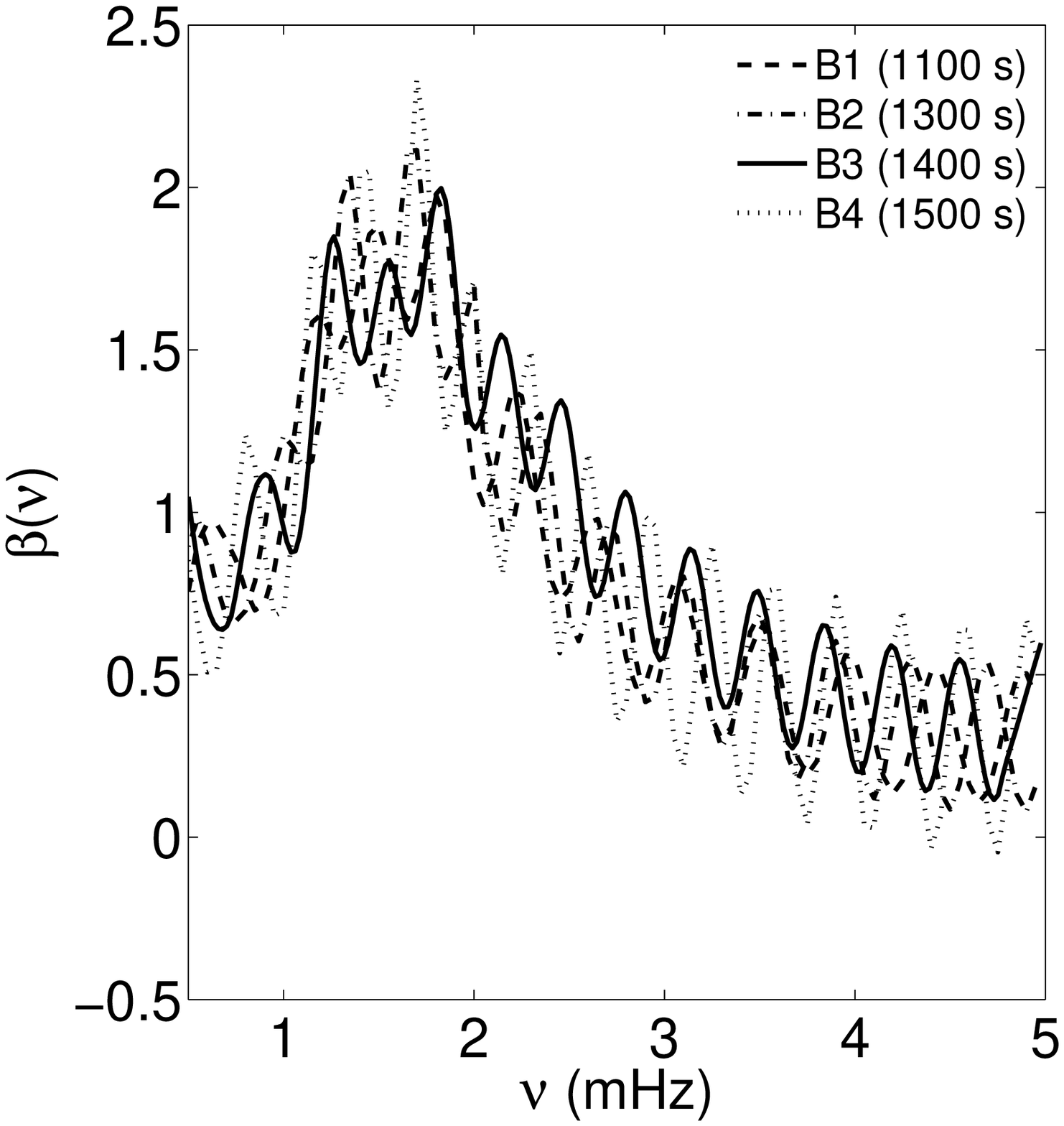}{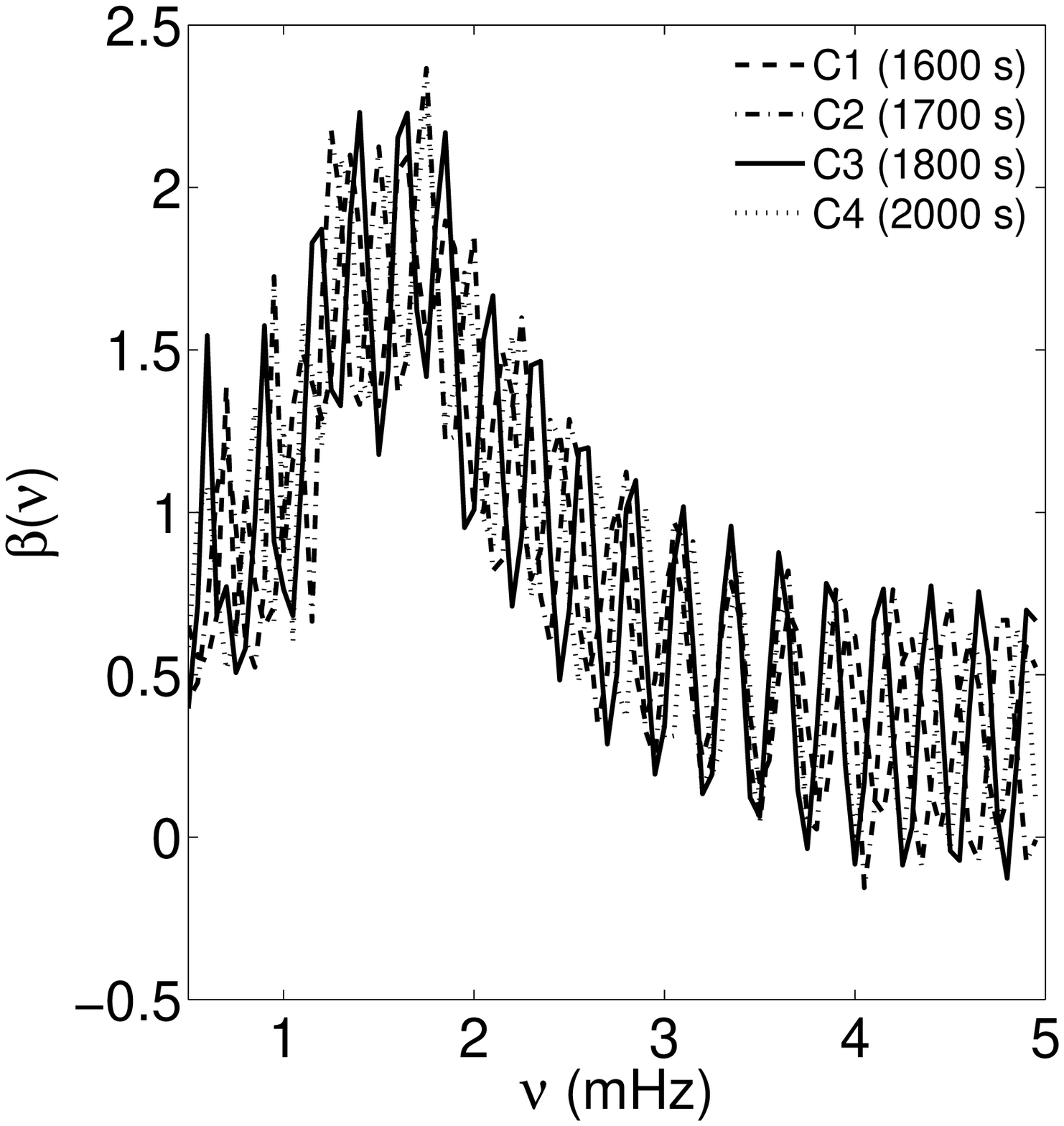}
\caption{Top left: several scenarios for a modified acoustic potential. A discontinuity in the first derivative of the sound speed was introduced to produce the glitches. We consider an interval in the acoustic depth from $\tau = 500$ s to $\tau = 2000$ s. The implications for the seismic parameter $\beta$ of these discontinuities are represented in the other three figures. The  ``period" of the feature decreases as it approaches the base of the convection zone. }
\end{figure*}

The fourth-order system of ordinary differential equations describing the linear adiabatic oscillations in the Cowling approximation (the gravitational effects are small in the outer layers) can be reduced to a second-order Schr\"odinger-type equation, considering that the trajectories of the acoustic waves in the surface layers are vertical \citep[e.g.,][] {1989nos..book.....U, 1989ASPRv...7....1V}. It follows that the equation of acoustic oscillations reads
\begin{equation}
\frac{d^2 \psi} {d \tau^2} + (\omega^2 - U^2) \psi = 0 ,
\end{equation}

where $\tau$ is the acoustic depth, $U$ the acoustic potential, $\omega (\equiv 2\pi \nu)$ the cyclic frequency
and  $\psi$ the wave function. The acoustic depth is given by
\begin{equation}
\tau = \int_r^R dr/c,
\end{equation} 
where $c$ is the adiabatic sound speed, and $r$ is the stellar radius.
The acoustic potential $U$ is defined by 
\begin{equation}
U^2(\tau) = \frac{g}{c} \left( \frac{g}{c} - \frac{d \ln h}{d \tau} \right) + \frac{1}{4} \left(\frac{d \ln \zeta}{d\tau} \right)^2 - \frac{1}{2} \frac{{d^2 \ln \zeta}}{d\tau^2} 
\end{equation}
with
\begin{equation}
h(r) = \rho^{-1}\exp \left(-2 \int_0^r \frac{g}{c^2} dr \right)
\end{equation}
and
\begin{equation}
\zeta(r) = \frac{r^2 h}{c},
\end{equation}
where $g$  is the radial gravity acceleration.  The method of the phase function \citep{1967SvPhU..10..271B, 1989ASPRv...7....1V, 1991SvA....35..400V} was used to represent the eigenfunction $\psi$
\begin{equation}
\psi(\tau, \omega) = A(\tau, \omega) \cos \theta,
\end{equation}
where $\theta =  \pi \alpha(\tau, \omega) + \pi / 4 - \omega \tau $, 
with an additional condition on the amplitude function
\begin{equation}
\frac{d\psi}{d\tau} = - \omega A(\tau, \omega) \sin \theta.
\end{equation}
The outer phase shift of acoustic waves $\alpha$ is computed  from the stellar envelope,
by numerically solving the following equation 
\begin{equation}
\frac{d\theta}{d\tau}=-\omega+\omega^{-1}U^2(\tau) \cos^2{\theta}.
\end{equation}

This equation is solved numerically
with the proper initial boundary condition at $\tau=0$.
The details of the numerical procedure can be found in~\citet{2001MNRAS.322..473L}.
$\beta (\nu)$  is obtained from numerical differentiation of $\alpha / \nu$ as
\begin{equation}
\beta(\omega) = \beta(\nu) = - \nu^2 \frac{d}{d\nu} \left(\frac{\alpha}{\nu} \right).
\end{equation}
The properties of the observable $\beta (\nu)$ were largely discussed by \citet{1988IAUS..123..137B, 1989SvAL...15...27B, 1991PASJ...43..739V, 1991SoPh..133..149M, 1994MNRAS.268..880R, 2001MNRAS.322...85R}, among others. These techniques, based on an analysis of the phase shift of acoustic waves, led to the measurement of solar helium abundance by \citet{1991LNP...388...43C} and \citet{1991Natur.349...49V} and to a contribution of the same authors to the calibration of the equation of state \citep{1992MNRAS.257...62C, 1992MNRAS.257...32V}.
The acoustic potential $U$ used in this calculation was
determined based on a detailed stellar structure model of a Sun-like star
specifically computed for this analysis.
The stellar evolution code CESAM~\citep{1997A&AS..124..597M} was used to compute this structure model. This code has an up-to-date and refined microscopic physics 
(updated equation of state, opacities, nuclear reactions rates,  and an accurate treatment of the microscopic 
diffusion of heavy elements), including the solar mixture of~\citet{2005ASPC..336...25A}. 
The basic input physics considered are identical to the solar standard model, as 
discussed in~\citet{1993ApJ...408..347T} and \citet{2013arXiv1302.2791L}. 
Figure 1 shows the value of $\beta(\nu)$ computed using  this method. The solid line represents the model being studied in this work, a star with 1.1 $M_{\odot}$. In this figure we have also included the parameter $\beta(\nu)$ calculated for a set of Sun-like stars with masses from $0.8\, M_{\odot}$ to $1.2\, M_{\odot}$.

\begin{table}
\caption{The ``periods" of the induced oscillations in the seismic parameter $\beta$ and the locations of the respective glitches}
\begin{center}
\resizebox{2.5cm}{!}{
    \begin{tabular}{ | c | c |}
    \hline
    
$  \tau (s) $    &  $\nu$ (mHz) \\ \hline
      500          &       0.800        \\\hline
      600          &       0.575        \\\hline
      750          &       0.619        \\\hline
      900          &       0.500        \\\hline
      1100          &     0.417        \\\hline       
      1300          &     0.350        \\\hline             
      1400          &     0.326        \\\hline             
      1500          &     0.310        \\\hline            
      1600          &     0.280        \\\hline             
      1700          &     0.260        \\\hline             
      1800          &     0.242       \\\hline             
      2000          &     0.233        \\\hline             
    \end{tabular}
 } 
\end{center}
\label{table: fits}
\end{table}

\subsection{Computing the Proxy of the Phase Shift from a Frequency Table}

\begin{figure}
\includegraphics[width=0.5\textwidth]{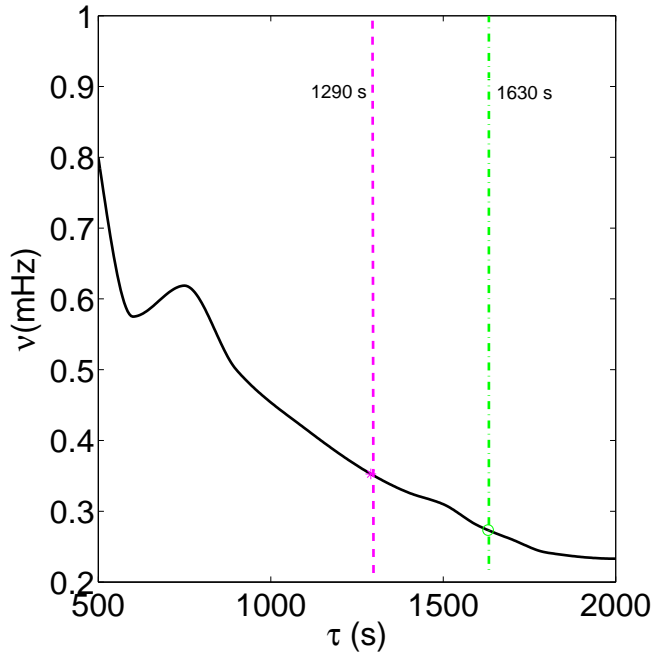}
\caption{Variation of the ``period"  of the oscillatory behavior of the phase $\beta$ with the location of an abrupt variation of the sound speed. This location is expressed in terms of the acoustic depth. The deeper the glitch, the higher the frequency of the oscillation induced in the phase $\beta$ by the glitch. The influence of the second ionization zone of helium around the $750$ s is visible. In this region, we have a slight increase of the frequency which contradicts the general tendency. Vertical bars show the location of a pretended glitch to induce an oscillation with the same period that we obtained from the observational $\beta$. The dashed line corresponds to the modes with degree $l=1$ (magenta line) and the dot-dashed line corresponds to modes with $l=2$ (green line).   }
\end{figure}

The seismic parameter $\beta(\nu)$ is the best proxy of the phase shift of acoustic waves
that  occurs  as a consequence of the scattering of these waves in the  upper layers of the star.
 $\beta(\nu)$,  as previously mentioned,  can be computed from a table (theoretical or observational) 
of frequencies. The  algorithm reads
\begin{equation}
\beta_{l,n} (\nu) = \frac{\nu_{l,n} - L \frac{\partial \nu}{\partial L} - n \frac{\partial \nu}{\partial n}}{ \frac{\partial \nu}{\partial n}},
\end{equation}

\begin{figure*}
\centering
\plottwo{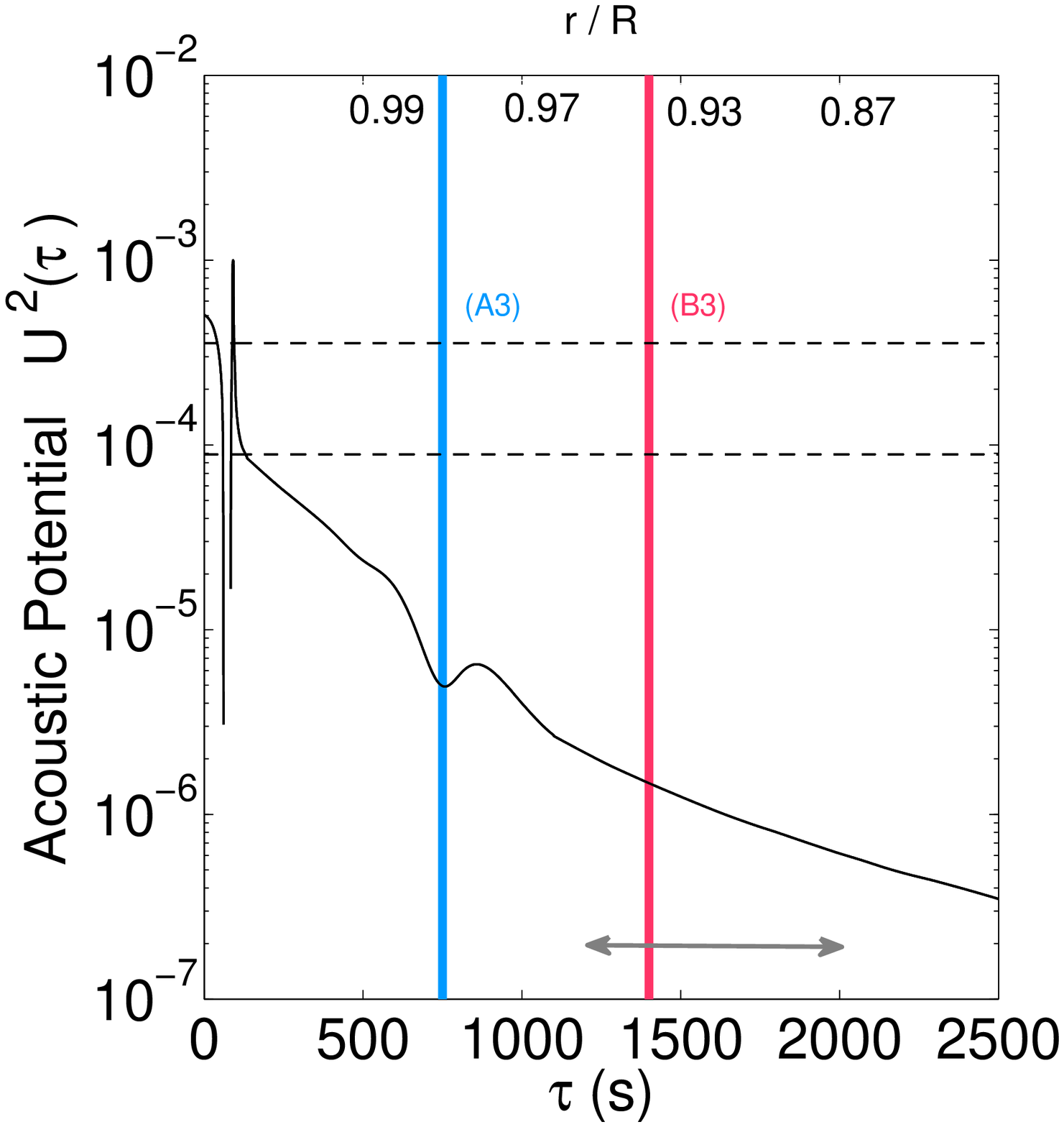}{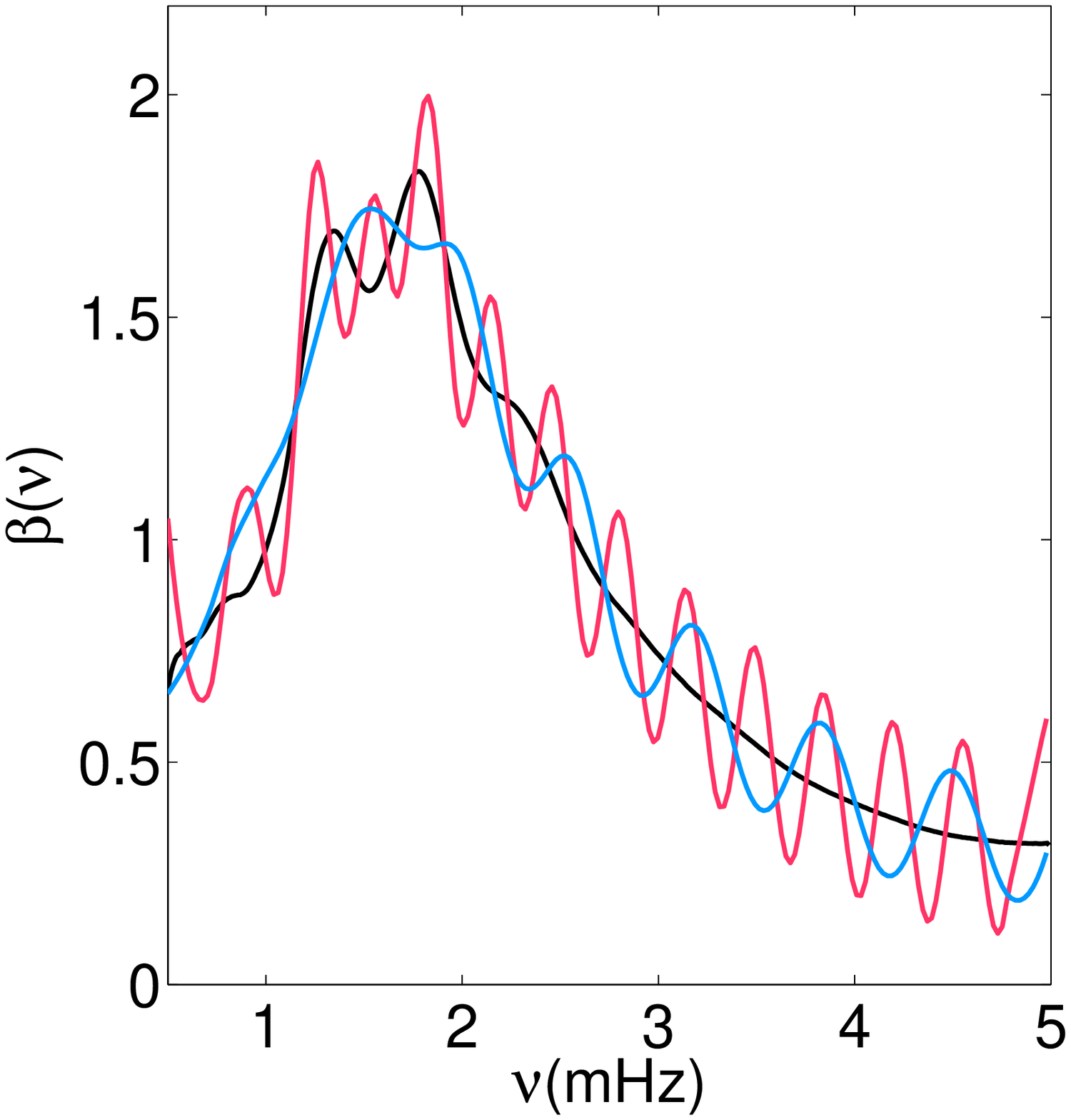}
\caption{(a) Acoustic potential obtained from the stellar envelope of a Sun-like star of $1.1\,M_\odot$. We can notice the superadiabatic region at $\tau \sim 80$\,s ($ \sim 0.9931R$ ), the zone of partial ionization of helium at $\tau \sim 750$\,s ($\sim 0.9840R$). The base of the convective zone is at $\tau \sim 2800$\,s ($\sim 0.7759R$). The horizontal dashed lines correspond to waves with frequencies $\nu = 1.5$ mHz and $\nu=3.0$ mHz. The vertical bars L1 (blue) and L2 (red)  represent the locations of two of our hypotheses for the existence of a rapid variation layer. The first one, L1, is located around  $\tau = 750$\,s whereas the latter, L2, is at an approximate acoustic depth of $\tau = 1400$\,s. (b) The seismic parameter $\beta(\nu)$ is in three cases: $\beta$ obtained from the ``unperturbed" acoustic potential (bell shape--black line); $\beta$ with a perturbation in the zone of the partial ionization of helium (oscillatory character and longer period--blue line); and finally, the $\beta$ with a perturbation located around the 1400\,s (oscillatory character and shorter period--red line).}
\end{figure*}

where $L = l + 1/2$ and $l$ and $n$ are the degree and radial order of the oscillation 
mode. We developed an algorithm that allows the calculation of these derivatives using the maximum of points available. This algorithm is based upon the formula proposed by \citet{1989SvAL...15...27B},
to study the acoustic phase shift in the Sun. The partial derivatives in $\beta_{l,n} (\nu) $  should be treated  carefully due to the small number of points available usually in observational frequency tables. In the specific case of this work we decided to use only the modes with degrees $l=1$ and $l=2$. This is due to the fact that the modes with degrees $l=1$ and $l=2$ are the modes with the biggest number of observational frequencies available, and thus, the modes that lead to a more reliable result.  The richness and properties of the observable $\beta (\nu)$ were largely discussed by Lopes and collaborators \citep{1994A&A...290..845L,1997ApJ...480..794L}.   $\beta_{l,n} (\nu) $ computed as indicated in Equation (10)  allows a direct comparison between the star and its stellar model, specifically,  highlighting the  
physical process differences in the upper layers of the model and the star.   
We computed $\beta_{l,n} (\nu) $ using a theoretical table for the same stellar structure model
used in the computation of $\beta(\nu)$ from the stellar envelope (see the previous section). 
The frequencies were obtained using a pulsation code that computes the adiabatic non-radial oscillations  
of the stellar structure model~\citep{2008Ap&SS.316..113C}. The theoretical table of frequencies obtained comprises around $83$ frequencies that correspond to acoustic modes with degree $l=1, 2$, and  $n$, taking values between $1$ and $25$. Figure 1 (b) shows the computation of the $\beta$ in the case of a Sun-like star with a mass of $1.1 \, M_\odot$.
Two methods of computing  $\beta$  are represented; the two theoretical descriptions of $\beta_{l,n} (\nu) $.
The computation of $\beta_{l,n}(\nu)$ from the theoretical frequencies  were performed over a range of frequencies between 1\,mHz and 5\,mHz.  The $\beta(\nu)$ computed from the stellar envelope has a very good overall agreement with the $\beta_{l,n}(\nu)$ calculated from the theoretical table of frequencies.  The general form of $\beta$ is very identical to the one obtained in the case of the Sun \citep{1989SvAL...15...27B}. However, $\beta(\nu)$,  in the case of the Sun spreads throughtout a range of frequencies 
between $1$ to $5$ mHz. In the case of this Sun-like star, the equivalent range is between $1$ to $3$ mHz. 
The shape of $\beta(\nu)$ is especially influenced by two factors: the superadiabatic region with a contribution of the hydrogen ionization zone, which defines the parabola form, and the zone of the partial ionization of helium which is responsible for the production of a sinusoidal component. This helium signature is evidenced in Figure 1, the larger the mass of the star the stronger the sinusoidal component. $\beta(\nu) $ also has a weak dependence on the initial phase.  Furthermore, in the case of low degree acoustic modes, it is known that there is a strong dependence of
$\beta_{l,n}(\nu)$ on the degree of the mode. This effect occurs due to the contribution of Eulerian perturbation of the
gravitational field to define eigen-frequencies of oscillation. This leads to $\beta_{l,n}(\nu) $ presenting a dependence with the degree $l$. This effect is very visible in the case
of the Sun \citep{2001MNRAS.322..473L}. This Sun-like star also shows the same effect. However, it is less visible (see Figure 1). This, in part, explains the difference between  $\beta$ computed from the stellar envelope (the solid line) and $\beta$ computed from the theoretical table of frequencies for $l=1,2$ (dotted lines).

\subsection{The Impact of a Rapid Sound Speed Variation in the Acoustic Potential}

\begin{figure*}
\centering
\plottwo{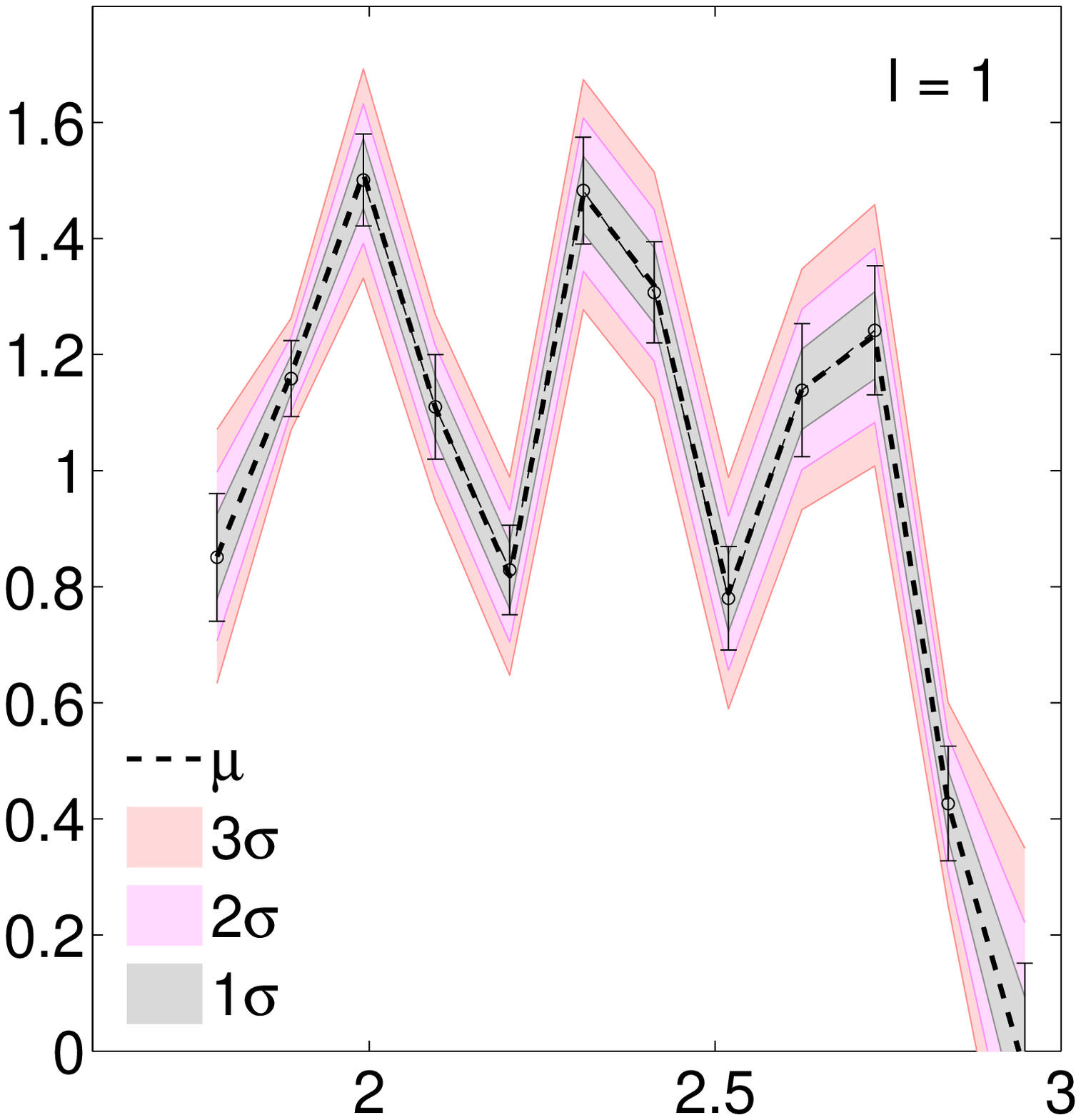}{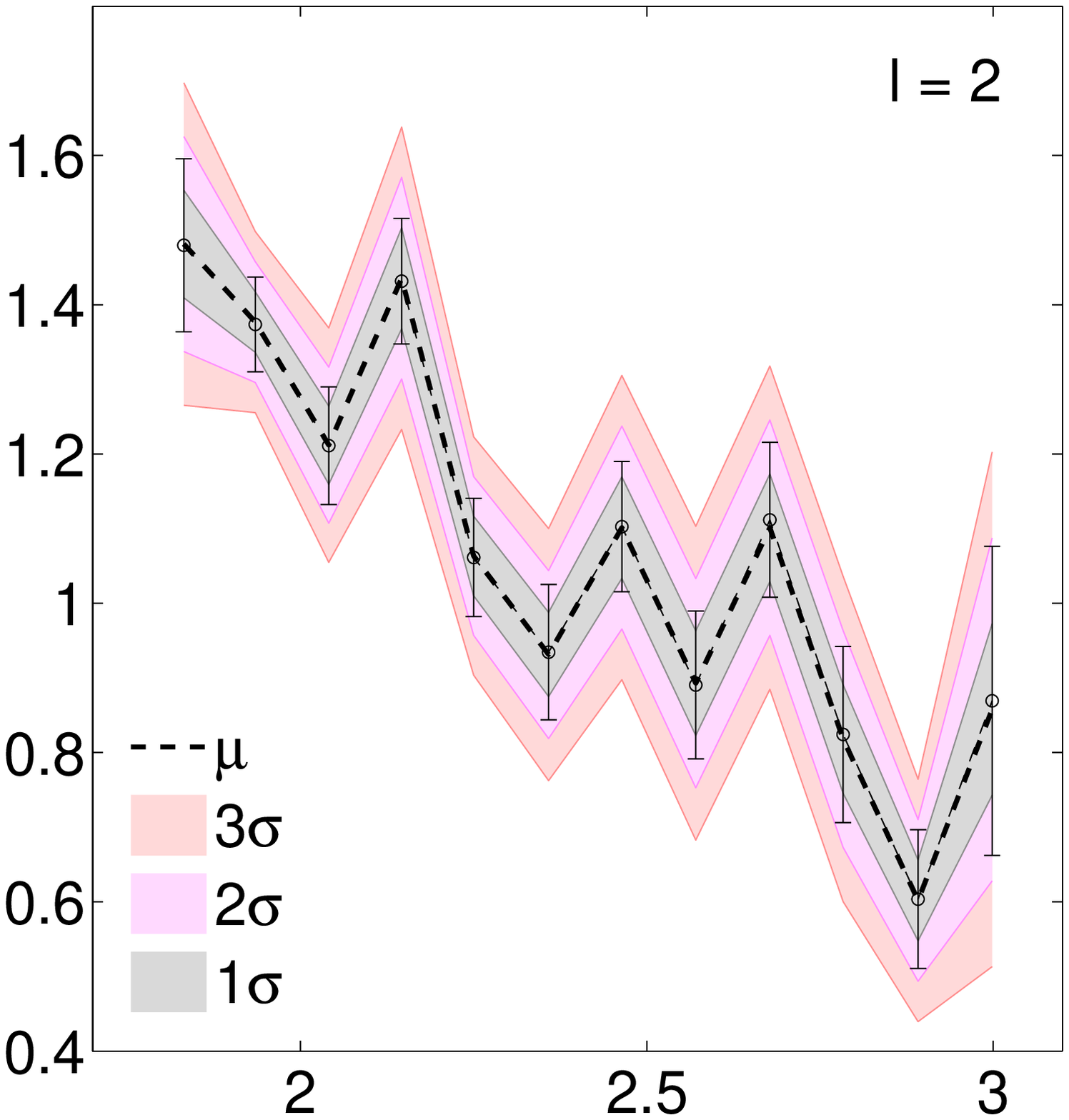}
\caption{$\beta_{\nu, l}$ parameter calculated for the observed frequencies of $\alpha$ Centauri A of the low degree modes $l=1,2$. The black dots represent the parameter $\beta_{\nu, l}$ for the observed frequencies with the error bars calculated for the respective values of the error measurements. Shadowed regions confirm the stability of the algorithm: they are obtained by adding white noise on the observed frequencies, $\nu \, ' = \nu \pm \Delta \nu$, where $\Delta \nu_{\text{max}} = 0.6 \, \mu\text{Hz} $.}
\end{figure*}

In the internal structure of stars, there are some regions where the sound speed changes rapidly due to changes in the structure. These regions are usually called glitches and they introduce an oscillatory behavior in the frequencies proportional to $\sin(2\tau_g\omega + \phi)$ where $\tau_g$ gives the location of the glitch \citep{1990LNP...367..283G}. As an example, in the Sun, the base of the convective zone and the helium ionization zone and their oscillatory signatures were studied in detail by  \cite{1994MNRAS.267..209B},  \cite{1994A&A...283..247M} and \citet{1994MNRAS.268..880R}. Application to Sun-like stars was a natural step forward: methods to locate the base of the convective zone and the He II ionization zone were studied and developed by  \citealt{2004A&A...423.1051B}, \citealt{2004MNRAS.350..277B}, \citealt{2007MNRAS.375..861H}, \citealt{2009arXiv0911.5044H}, \citealt{2000MNRAS.316..165M}, \citealt{2001A&A...377..192M}, \citealt{2005A&A...441.1079M} and \citealt{2003A&A...411..215R}. All these regions are in fact transition layers, where a sharp variation of the sound speed occurs. Acoustic sound waves are reflected and refracted in the surface of the star and by these layers. The phase shift generated by these regions was estimated by \citet{1994MNRAS.268..880R}. Considering that the acoustic potential has a variation of $\delta U^2 = U^2_2 - U^2_1$ within an interval $\tau_1 < \tau < \tau_2$, there will be a variation of the eigenfuncion $\delta \psi = \psi_2 - \psi_1$. From Equation (1) we can reach the following Wronskian
\begin{equation}
[ W(\psi_1, \psi_2)]^{\tau_2}_{\tau_1} = \int_{\tau_1}^{\tau_2} \delta U^2 \psi_1\, \psi_2\, {d}\tau
\end{equation}
where
\begin{equation}
W(\psi_1, \psi_2) = \psi_1 \frac{d \psi_2}{d \tau} - \psi_2 \frac{d \psi_1}{d \tau}.
\end{equation}
The eigenfunction $\psi$ given by Equation (6) allows us to rewrite the Wronskian
\begin{equation}
[ W(\psi_1, \psi_2)]^{\tau_2}_{\tau_1} = \omega A_1(\tau_2, \omega) A_2(\tau_2, \omega) \sin(\pi \alpha_0).
\end{equation}
This is obtained considering the outer boundary condition at $\tau = 0$ ($r = R$) and $\alpha_0 = \alpha_2(\tau_2, \omega) - \alpha_1(\tau_2, \omega)$. At a first approximation, and not taking into account the variation of the amplitude for $\tau_1 < \tau < \tau_2$, we have
\begin{equation}
\alpha_0 \cong \frac{1}{\pi \omega} \int_{\tau_1}^{\tau_2} \delta U^2 \cos^2 [\pi \alpha(\tau, \omega) + \pi / 4 - \omega \tau] d \tau,
\end{equation}

where $\alpha(\tau, \omega)$ is the phase shift of the unperturbed eigenfunction. 
This is a general expression which approximately estimates the phase shift $\alpha_0(\nu)$ induced by a variation of the acoustic potential within an interval in the acoustic depth for some value of the frequency $\omega$. As we can see, this induced phase shift will depend on the particular feature that is described by $\delta U^2$. 
Discontinuities in the first derivative of the sound speed produce glitches in the acoustic potential and can be represented by a Dirac $\delta$ function. We will see that this perturbation in the potential will produce an additional periodic component. Indeed, if we represent the acoustic potential as
\begin{equation}
U^2(\tau) = U_0^2(\tau) + a_g \delta(\tau - \tau_g)
\end{equation}
where $U_0^2$ is the unperturbed potential, $a_g$ is the contribution to the acoustic potential from the glitch (i.e from the discontinuity in the first derivative of the sound speed) and $\delta$ is the Dirac $\delta$-function, we obtain directly from integral (14)
\begin{equation}
\alpha_g \cong \frac{a_g}{\pi \omega} \cos^2 \left[\pi \alpha(\tau_g, \omega) + \frac{\pi}{4} - \omega \tau_g\ \right].
\end{equation}

This $\alpha_g$ can be seen as the quasi-periodic contribution to the phase shift which originates from the discontinuity in the first derivative of the sound speed.
We numerically simulated a small variation of sound speed, typically smaller than $1\%$, and realized that small variations of sound speed of such magnitudes can originate acoustical glitches in the potential. 
In this procedure, several different locations for $\tau_g$ were tested,
namely between the  acoustic depth of $\tau = 500$\,s (0.99$R$) and  $\tau = 2000$\,s (0.88$R$). Figure 2 illustrates these tests made with the modified acoustic potential and the correspondent effects on the seismic parameter $\beta(\nu)$. The ``period" of the observed quasi-periodic signal is determined by the location of the RV region and its amplitude is determined by the magnitude of the discontinuity in the acoustic potential \citep{1993ASPC...42..169R}. Table 1 shows the different locations tested and the corresponding periods of the oscillation generated by the RV region. In Figure 3, we fitted the values of Table 1 with a Piecewise Cubic Hermite Interpolating Polynomial. We clearly see the decreasing tendency of the period of the oscillatory feature, except in the He II ionization region. 
In particular, we want to focus now on the two cases of Figure 4. This figure shows two of the previous scenarios: the A3 (blue) and the B3 (red) layers represent the positions where the glitch is located. Figure 4(b)  shows  $\beta(\nu)$ computed for these two new scenarios,  as well as $\beta(\nu)$  corresponding to the non-modified potential (see Figure 1). A detailed analysis of the observed $\beta$ suggests that the real $\alpha$ Centauri A could have an identical potential feature located in the
external layers of the star.
However, a recent study by \citet{2011MNRAS.414.1158C} of the overshoot region at the bottom of Sun's convective zone, suggests that the regions of transition in real stars might be smoother than generally predicted by theoretical models.

\section{Toy model: the acoustic scattering of $\alpha$ Centauri A}

 \subsection{Fundamental Aspects of the Star $\alpha$ Centauri A}

The stars of the $\alpha$ Centauri stellar binary system are among the most studied and observed 
Sun-like stars. Being this  binary system the closest to the Earth, it makes it so its stars  
have a well known set of observational constraints.  This work was applied to the study of the upper layers of $\alpha$ Centauri A, the most massive star in this binary system. This was done using the acoustic phase shift and some seismic properties of this star.
 A large number of stellar evolution models have been developed for this star. The first  model was computed by \citet{1978ApJ...221..175F}, where they assumed a solar composition for the star $\alpha$ Cen A, but  were not able to obtain the precise astrometric and photometric parameters. Until the year 2000, several models were proposed, based exclusively on non-asteroseismic constraints: \citet{1991A&A...247...91N}, \citet{1992ApJ...394..313E}, \citet{1993A&A...268..650N}, \citet{1993ApJ...403L..79L}, \citet{1995A&A...295..678F}, \citet{2000A&A...363..675M}, and \citet{2000ApJ...531..503G}. All these theoretical models were unable to give an unambiguous description of the stellar parameters.
 In 2002, \citet{2002A&A...390..205B}, obtained a definitive identification of {\it{p}}-mode observational frequencies. In the same year, \citet{2002A&A...386..280P} computed, with improved precision, the masses of the $\alpha$ Centauri system. Their work was based on the accurate estimate of the parallax by \citet{1999A&A...341..121S}. One year later, \citet{2003A&A...404.1087K} using  new interferometric measurements made for this star, provided accurate measurements of the angular diameters for the $\alpha$ Cen system, which allowed the precise determination of the radii. These new constraints generated a new set of models for the stars $\alpha$ Centauri: \citet{2002A&A...392L...9T}, \citet{2003A&A...402..293T}, \citet{2004A&A...417..235E}, and \citet{2005A&A...441..615M}. These models were able to fit some of the asteroseismic constraints, but they left some questions unanswered, such as the nature of the core of the star. Our $\alpha$ Centari A model was calibrated to have the present stellar radius 
$ R_\star= 8.519 \times 10^{10} \;{\rm cm}$,  luminosity  $L_\star = 5.816 \times 10^{33} \; {\rm erg\; s}^{-1}$, 
mass $M_\star = 2.198 \times 10^{33} \;{\rm g}$ \citep{2003A&A...404.1087K, 2004A&A...417..235E, 2002A&A...386..280P} and age $t_\star =5.8 \; {\rm Gyr}$. 
The model was also required to have a fixed value of the photospheric ratio $(Z/X)_\star$, where  
 {\it{X}} and {\it{Z}} are, respectively, the mass fraction of hydrogen and the mass fraction of elements heavier than helium.
The value of $(Z/X)_\star$ we adopted was proposed by \citet{2003A&A...402..293T}.
Our model is identical to other models of $\alpha$ Centauri A found in the literature. It is specifically similar to the model A3 with a radiative core from \citet{2005A&A...441..615M}.

Recently, \citet{2010A&A...523A..54D} published an observational table of frequencies that unified the data from CORALIE \citep{2002A&A...390..205B} and from UVES and UCLES \citep{2004ApJ...614..380B}. The goal was to compute a combined velocity time series in order to reduce the daily aliases in the power spectrum. They were able to present a frequency table in which
 they identify 44 acoustic modes. These 44 acoustic modes allow a reliable seismic study of the acoustic phase shift, especially for the modes with degrees $l=1,2$.

\begin{figure}
\includegraphics[width=0.5\textwidth]{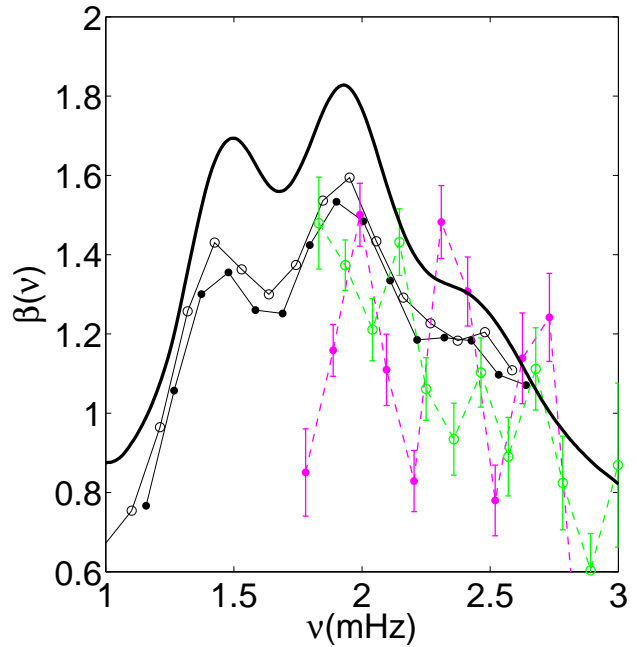}
\caption{Theoretical and the observational seismic signature $\beta$. The continuous curves are the theoretical curves ($\bullet$  $l=1$, o  $l=2$) and the dashed curves represent the $\beta$ calculated from an observational frequency table for the modes with the degrees ($\bullet$  $l=1$, o  $l=2$). }
\end{figure}

\subsection{Seismic Observable $\beta$}

\begin{figure*}
\centering
\plottwo{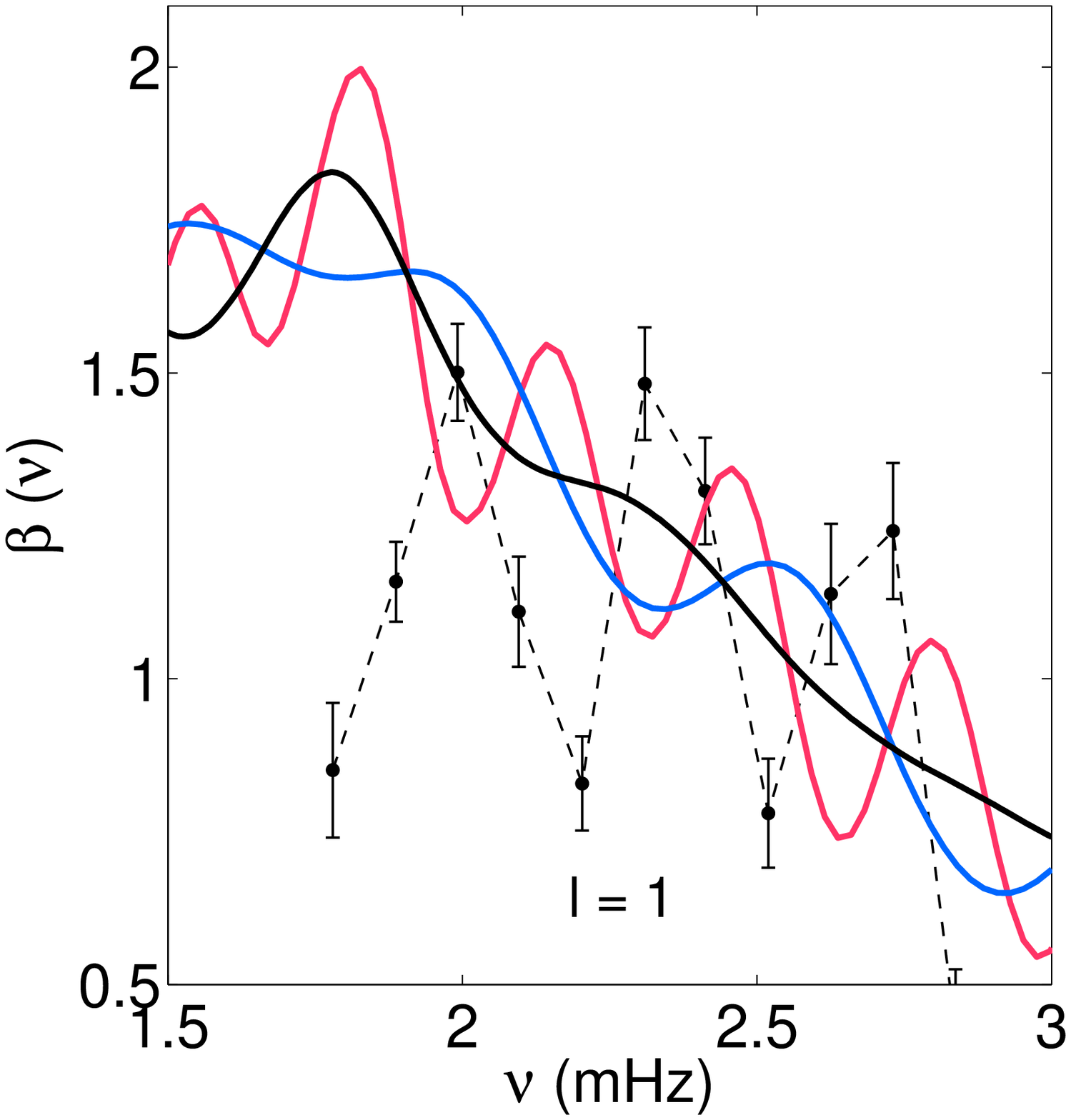}{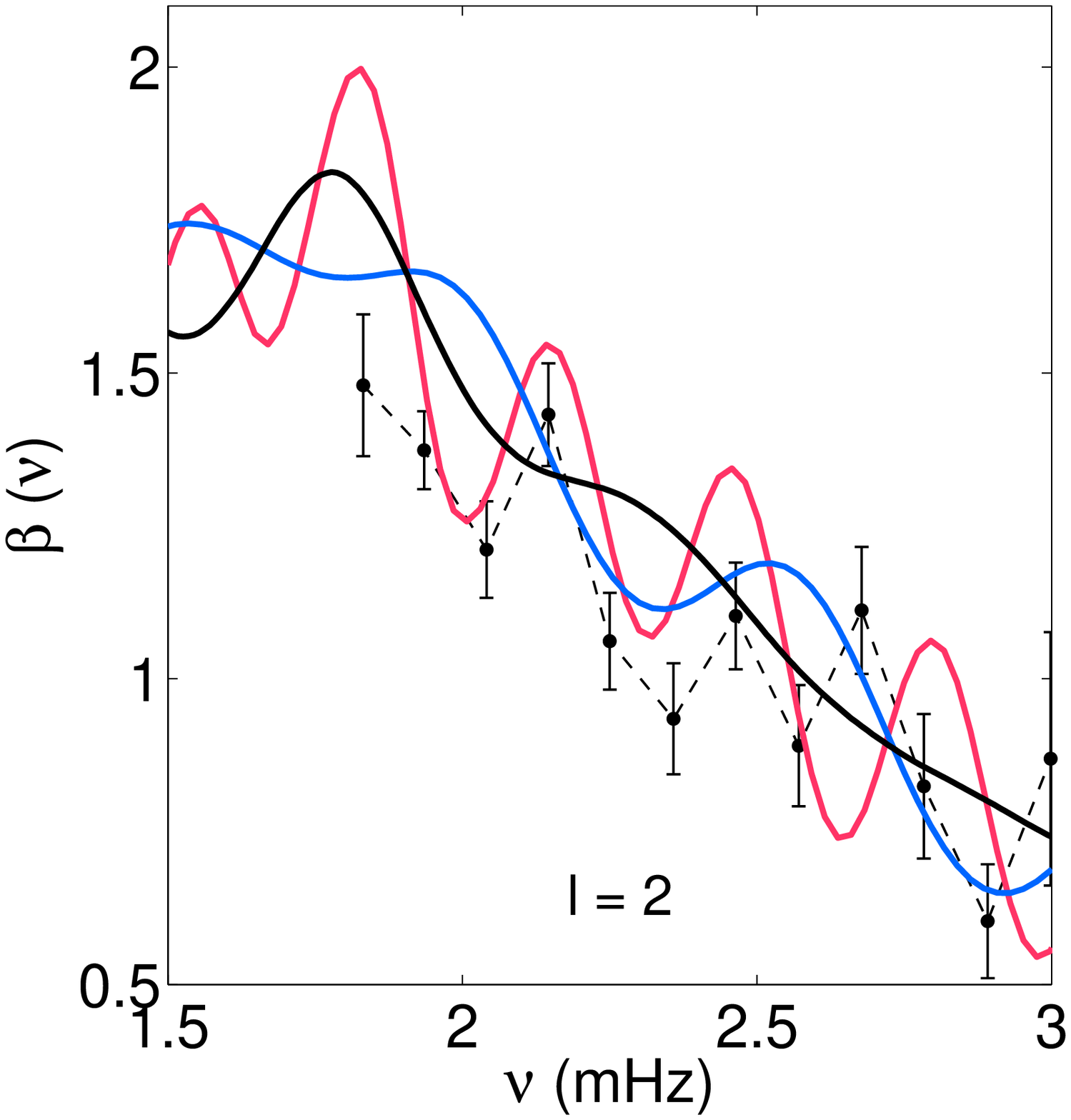}
\caption{Theoretical seismic parameter $\beta$ obtained from modified and non-modified acoustic potentials, compared with the observational $\beta$ for the star $\alpha$ Centauri A. Black solid line: theoretical $\beta$ from an unmodified acoustic potential. Blue line: theoretical $\beta$ from a modified acoustic potential with a glitch in the He II ionization zone. Red line: theoretical $\beta$ from a modified acoustic potential with a glitch around an acoustical depth of 1400 s. Dashed black line: observational $\beta$ calculated from the table of frequencies of  \citet{2010A&A...523A..54D}.\\ }
\end{figure*}

Our aim now is to recover $\beta_{l,n}$, using Equation (10) from the observational table of frequencies of the star $\alpha$ Centauri A.
We used the frequency table of \citet{2010A&A...523A..54D} which is so far the most updated table for this star. The derivatives in Equation (10) were calculated using up to five-point central differences formulas. The problem with any numerical differentiation is directly connected with the structure of the experimental data: small perturbations on the function to be differentiated may cause large errors in the numerical derivatives. In our case, the frequencies used to evaluate the derivatives differ by a value of $\sim 105\, \mu \text{Hz}$ and the uncertainties in the frequencies are, in the worse case, given by  $\sim 0.6\, \mu \text{Hz}$. This means that the error in calculating the derivatives will never exceed  a few percents. Several tests were performed to verify the quality and stability  of the algorithm 
to compute $\beta$  from this observational frequency table. In particular, we have repeatedly added a random white noise on the observed frequencies ($\nu\, ' = \nu \pm \Delta \nu$). This white noise was added within the larger value of the frequency measurement error ($0.6\, \mu \text{Hz}$) and the procedure was executed 400 times. At the same time, we used the propagation error theory to calculate the error bars associated with the exact value of the uncertainties of the observational frequencies. The results are presented in Figure 5 for $l=1,2$ over a range of frequencies from $\sim$ 1.8\,mHz to $\sim$ 3\,mHz and confirm that the algorithm is stable.
In Figure 6 we superimposed the theoretical signatures with the observational $\beta$.
We can notice a dispersion in the observational modes that is not present in the theoretical modes. This dispersion, as we will see, carries information about the oscillatory character present in each set of individual acoustic modes with the same degree. The possibility of using this observable as a diagnostic tool and its limitations in the case of Sun-like stars was discussed by \citet{1997ApJ...480..794L}.
A detailed analysis of the observed $\beta$ suggests that the real star could have an identical potential feature located in the
external layers of the star. To illustrate the origin of the oscillatory character of the observational $\beta_{l,n}$, we compare
the  $\beta (\nu)$ of the (B3) scenario (red) in Figure 3 with $\beta_{l,n}$ computed from the observational frequency
table, for which the $\beta_{l,n}$ of acoustic modes with the same degree are connected with a single line (see Figure 7). We also include in this comparison the (A3) scenario (blue) and the $\beta(\nu)$ calculated from the unmodified acoustic potential, the black line in Figure 4.
These tests with the modified acoustic potential are able to reproduce the oscillatory behavior of $\beta_{l,n}$ 
(set of frequencies  with the same $l$ are connected with a single dashed line) better than the unmodified potential. For the "period" of the observable $\beta$ we estimated an error of $\sigma_P \le \sqrt{2} \sigma_{{\text{max}}} \approx 0.9\, \mu\text{Hz}$ using error propagation theory. This means that for $l=1$ we will have an error of $\sim 0.2\%$ and for $l=2$ this same error is $\sim 0.3\%$ in the respective ``periods".
We found that the location of the {\it RV layer} that best fits the observational
$\beta_{l,n}$(with fix $l$) is located  around $\tau = 1400$\,s (0.94 $R$).
This feature is not present in the unmodified $\beta_{l,n}$ computed from the theoretical table and is not predicted by theoretical models. \citet{2004A&A...423.1051B} had noticed the presence of an oscillation corresponding to an acoustic depth of $\sim 1300\,\text{s}$ in this star, but considered it as an artifact.

The range of acoustic depths available to perform a study with the characteristics above is determined by the value of the large separation of the star. We should be aware if the order of magnitude of the separation of the points, used to evaluate the observable $\beta$ is similar to the period we want to determine. Indeed, in this case, the ability to measure the period of the signature $\beta$ will be lost, since we have reached the limit of resolution of the method which can be expressed by the sampling theorem (Kotelnikov/Nyquist-Shannon). The values of Table 1 together with Figures 2, 4(b) and 5 show that the deeper we move inside the star the closer we are to the resolution limit.

\begin{table*}
\caption{Coefficients of the sinusoidal fit to the observational oscillation of $\Delta \beta(\nu)$}
\begin{center}
\resizebox{16cm}{!}{
    \begin{tabular}{ | c | c | c | c | c | c | c |}
    \hline
    \multicolumn{7}{| c |}{ \bf{  $ f(\nu) =  A\sin(\omega_0 \nu+ \delta)$}} \\
    \hline
    \hline
    $l$             & A                                       & $\omega_0$              & $\delta$                   &  {\it{T}}   (mHz)            &     $\tau$ (s)            &        $ r/R$         \\

  $  1 $          &   0.3318 $\pm$ 0.1891     &  18.11 $\pm$ 1.970   & 9.475 $\pm$ 4.380 &    0.347 $\pm$ 0.0378    & 1440 $\pm$ 314    & $0.936^{+0.027}_{-0.035}$    \\
$  2 $            &    0.1584  $\pm$ 0.0968   & 19.87  $\pm$  2.780  & 9.097 $\pm$ 6.213 &   0.316 $\pm$ 0.0443     & 1580 $\pm$ 443    &$0.921^{+0.041}_{-0.058}$       \\ \hline

    \end{tabular}
 } 
\end{center}
\label{table: fits}
\end{table*}

\subsection{Fitting a Sinusoidal Wave to Data}

When an abrupt change occurs in the internal stratification of a star, the frequencies of oscillation will show a characteristic periodic signal. These are regions of RV of the sound speed, and the observable seismic parameters are consequently affected by the oscillations in the frequencies. The most studied example of such a region is the base of the convective zone in the Sun. The RV of the sound speed associated with the base of the convective zone in the Sun was studied in detail by several authors in the past \citep{1990LNP...367..283G, 1994A&A...283..247M, 1994MNRAS.268..880R, 1997MNRAS.288..572B, 2001A&A...368L...8M, 2004A&A...423.1051B}. Specifically, it was found that the location of the convective region can be extracted from the frequencies. \cite{1990LNP...367..283G}  was the first to associate the period of such an oscillation with the acoustic depth of the glitch. \cite{1994A&A...283..247M, 2000MNRAS.316..165M} although using a different approach also showed that these oscillations can be found in the large separation and in the second difference. Higher order differences were discussed by \cite{1997MNRAS.288..572B} and \cite{2001A&A...368L...8M}. In this work, we are mainly interested in the seismic parameter $\beta$. This phase-shift derivative was investigated in detail by \citet{1994MNRAS.268..880R, 1996MNRAS.278..940R, 2001MNRAS.322...85R}. The phase shift $\alpha(\nu)$ is known to vary quasi-periodically with the  frequency $\nu$ with a period  of $\approx 1/2\tau_g$ where $\tau_g$ is the acoustic depth of the location of the RV region \citep{1994MNRAS.268..880R}. There is no reason to think that these techniques cannot be applied to other Sun-like stars.
Based on the previously mentioned studies we made a test to the observational $\beta_{l,n}$ with the intent to isolate
this new oscillation feature. In this approach,  we start to compute the difference between the observational and the theoretical 
values of the seismic parameter $\beta$, i.e.  $\Delta \beta= \beta_{\text{obs}} - \beta_{\text{th}}$.
The $\beta_{\text{th}}$ was obtained by making a spline fit to the theoretical values of $\beta_{l,n}$ 
(see the black dots in Figure 1). We think this is a more accurate representation of the theoretical
 $\beta $ of the star rather  than using the  $\beta $  computed from the stellar envelope. 
This procedure  allows us  to eliminate any contribution from the superadiabatic region and the second helium ionization zone from the observations. Following such procedure, we performed a sinusoidal least-squares fit to each  $\Delta \beta$ with fixed $l$. The results are shown in Table 2 and the coefficients of the fits were determined with 95\% confidence intervals. Here again, if we use the {\it{period}} of each oscillation to estimate the location of the acoustic waves associated with {\it RV layer}, we find  the following results: for  $\Delta \beta$ computed for modes with degree $l=1$ and $l=2$ , the angular frequencies are $\omega_0 = 18.11 \pm1.970$ and   $\omega_0 = 19.87 \pm 2.780$ respectively, which implies the frequency will be $\nu=\omega_0 / 2\pi$. Next, we can calculate the ``period" $T= 1/\nu$ and this period gives us the acoustic depth $\tau \approx 1/2T$. These results suggest  the existence of the {\it RV layer}  below the region of partial ionization of helium. However, its exact location seems to be more difficult to determine.  

It is well known that the upper turning points for low frequency modes are located deeper in the surface layers than the turning points for modes with higher frequencies. Considering the observational frequencies, a typical mode of 2.4 mHz, will have its upper turning point at $r/R = 0.999$ whereas a mode with a lower frequency of 1.7 mHz has its upper turning point at $r/R = 0.992$. We took into account this shift and found no significative differences for the location of the glitch in terms of normalized radius $r/R$ (in the case of the mode with degree $l=1$ we found $r = 0.935\,R$ instead of $r=0.936\,R$ listed in Table 2 and no differences in the case of the mode with degree $l=2$).  We think it is possible to say that the radial location of the glitch in Table 2 is not affected by the vertical stratification in the surface layers and the upper reflection boundary is well approximated by the photospheric radius.
 All the values are shown in Table 2 with the respective uncertainty. To have an idea of the expression of this uncertainty we represented all the range of this uncertainty with a grey double arrow in Figure 4.

Some seismic indicators frequently used to identify sharp variations of the sound speed in a Sun-like star, display an effect known as the aliasing problem. This effect states that for modes of pulsation with the same value of the degree $l$ it is not possible to distinguish from discontinuities located at an acoustic depth of $\tau$ or $\tau_0-\tau$. Here, $\tau_0\equiv\tau(0)$ is the total acoustic radius. This effect was first noticed by \citet{2001A&A...377..192M} and then by \citet{2003MNRAS.344..657M} in the case of gravity modes. We think that the method described in this paper, which uses the seismic observable $\beta$, is safe from this aliasing effect. By definition, $\beta_{l,n} (\nu)$ (see Equation (10)) is computed considering the maximum number of modes available (with $l=0,1,2,3$). Nevertheless, this method should, whenever possible, be used in combination with other seismic methods to validate the results.

\section{Summary and Conclusion}

In this work, we explore the diagnostic capabilities of ``the acoustic phase shift method'' to probe the detailed structure of stellar envelopes. We specifically develop a technique to infer the impact that a small  RV layer (located in the upper layers of the star) and a glitch of the sound speed have on the acoustic phase shift, or equivalently, on the frequencies of  the  acoustic modes. In particular, we test our method in the case of the spectrum of acoustic oscillations of a $1.1\,{M_\odot}$ Sun-like star. The diagnostic of the acoustic phase shift is obtained by means of a functional constructed from the table of frequencies, usually known as the seismic parameter $\beta (\nu)$. This observable $\beta$, that is a derivative of the phase of the acoustic modes, is therefore used  to deduce the properties of  the  physical processes occurring in the upper layers of a Sun-like star. 

Following that, we decided  to test our technique in the case of the star $\alpha$ Centauri A. We have computed $\beta(\nu)$ for the theoretical and observational tables of frequencies. We noticed that there is a disagreement between the theoretical $\beta$ (obtained numerically by integrating the wave equation through the external layers and from a theoretical frequency table) and the observational $\beta$. This disagreement is due to a sinusoidal-like feature found in the observational $\beta$, but absent in the theoretical one. Such surprising  result obtained in this preliminary study of  $\alpha$ Centauri A, motivated  us to inquire further about this observed feature and further test the validity of  our result by computing $\beta $ for several stellar envelopes and by comparing $\beta$ with other seismic diagnostics. Using $\beta(\nu)$ as a probe tool, we have constructed several stellar envelope models with a local perturbed sound speed profile. We found that only a small RV on the sound speed (or equivalently in the acoustic potential), i.e., a glitch,  could explain the origin of this observed sinusoidal-like feature in $\beta$. We also found that the period and amplitude of this sinusoidal-like feature gives the location, thickness, and height of the glitch. In this specific case of $\alpha$ Centauri A, the  glitch is located around an acoustical depth of $\tau \sim 1400$ s ($\sim$ 0.940$R$). 
The existence of  a  helium II ionization layer in the envelope of these types of stars, is among the physical processes that best explain such types of sinusoidal-like features in the stellar acoustic spectrum, or $\beta$, including the star used in our case study - $\alpha$ Centauri A. Indeed, the fact that a glitch located in  the envelope of  a Sun-like star could produce a sinusoidal-like signature in the acoustic modes was first pointed out by~\citet{2009arXiv0911.5044H}. We believe that  even if this is not the only way to obtain this type of sinusoidal like feature, this physical process is among the best candidates to explain it. In particular, in the case of $\alpha$ Centauri A,  this could explain the existence of the RV layer located 6\% below the surface of the star. However, stellar evolution models of these types of Sun-like stars, predict the location of  the helium II ionization around 2\% below the surface, which is well above the value found from the  observational data. The exact physical mechanism responsible for the scattering of the acoustic waves in the Sun-like stars is still unknown and needs further investigation. Nevertheless, there are a few indications from theoretical and observational studies that favor the existence of these rapid transition layers in the interior of  Sun-like stars. Among others, we could mention the partial ionization of heavy elements (for example Carbon, Nitrogen and Oxygen), which in some stars such as  sub dwarf B stars~\citep{2006ApJ...648..637R} excited oscillations by the opacity mechanism, near surface rotational shear layer connected with meridional flows within the upper part of convection zone~\citep{2009LRSP....6....1H}, and physical processes such as penetrative convection~\citep{2012A&A...544L..13L}. It is worth mentioning  that the existence of convective cores in stars, especially, if their mass is above  the critical mass value of $1.1\ M_\odot$~\citep{2013ApJ...769..141S}, could increase the amplitude of this observed sinusoidal-like  feature. Indeed, we made several tests and found that in the case of a Sun-like star with a convective core, it is still possible to observe a sinusoidal-like feature  in $\beta$, identical to the one observed  in Figures 3 and 5. It is worth pointing out that the period  associated with  this feature rules  out the possibility of  it being  created by the convective core. However, we think that  waves traveling from the core toward the surface, when perturbed by a RV layer of a glitch located in the stellar envelope, will experiment a wave interference process that could reinforce or even amplify the observed sinusoidal-like feature produced by the glitch.   

Finally, it is important to  highlight that this method, based on the scattering of the acoustic waves in the outer layers of the Sun-like stars, can also be used  to study others stars, and thus can contribute to and increase our knowledge of their interiors. The observable $\beta(\nu)$ is a natural theoretical-observable of the acoustic  phase shift  which can become a powerful tool to extract information on the stellar structure from the acoustic global modes.

\begin{acknowledgments}

We want to thank to the referee for his comments that led to a deepening of the theoretical aspects of the relation between a region of rapid variation  of the sound speed inside a star and its signature in the phase shift. Referee comments also resulted in a more accurate and robust manuscript.

We are grateful to P. Morel for making the CESAM code for stellar evolution available and 
to J. Christensen– Dalsgaard for his Aarhus adiabatic pulsation code (ADIPLS).

This work was supported by grants from ``Funda\c c\~ao para a Ci\^encia e Tecnologia" (SFRH/BD/74463/2010).
\end{acknowledgments}
%






\end{document}